\documentclass[preprintnumbers,amsmath,amssymb,floatfix,12pt,prd,superscriptaddress,nofootinbib]{revtex4}
\usepackage{graphicx}
\usepackage{epsfig}
\usepackage{bm}
\usepackage{amsfonts}   

\begin{document}

\title{Holographic dark energy in Brans-Dicke cosmology with Granda-Oliveros cut-off}

\author{Mubasher Jamil}
\email{mjamil@camp.nust.edu.pk} \affiliation{Center for Advanced
Mathematics and Physics, National University of Sciences and
Technology, H-12, Islamabad, Pakistan}

\author{K. Karami}
\email{KKarami@uok.ac.ir} \affiliation{Department of Physics,
University of Kurdistan, Pasdaran St., Sanandaj, Iran }
\affiliation{Research Institute for Astronomy $\&$ Astrophysics of
Maragha (RIAAM), Maragha, Iran}

\author{A. Sheykhi}
\email{sheykhi@mail.uk.ac.ir} \affiliation{Department of Physics,
Shahid Bahonar University, P.O. Box 76175, Kerman,
Iran}\affiliation{Research Institute for Astronomy $\&$
Astrophysics of Maragha (RIAAM), Maragha, Iran}

\author{E. Kazemi}\affiliation{Department of Physics, Shahid Bahonar University,
P.O. Box 76175, Kerman, Iran}

\author{Z. Azarmi}
\affiliation{Department of Physics, University of Kurdistan,
Pasdaran St., Sanandaj, Iran }

\date{\today}

\begin{abstract}
\vspace*{1.5cm} \centerline{\bf Abstract} \vspace*{1cm} Motivated by
the recent works of one of us \cite{Karami1,Karami2}, we study the
holographic dark energy in Brans-Dicke gravity with the
Granda-Oliveros cut-off proposed recently in literature. We find out
that when the present model is combined with Brans-Dicke field the
transition from normal state where $w_D
>-1 $ to the phantom regime where $w_D <-1 $ for the equation of
state of dark energy can be more easily achieved for than when
resort to the Einstein field equations is made. Furthermore, the
phantom crossing is more easily achieved when the matter and the
holographic dark energy undergo an exotic interaction. We also
calculate some relevant cosmological
parameters and their evolution.
\end{abstract}


\maketitle

\newpage
\section{Introduction}

Recent cosmological observations obtained from distant supernovae
SNe Ia {\cite{c1}}, cosmic microwave background radiation
explorations of WMAP {\cite{c2}}, galaxy redshift surveys of SDSS
{\cite{c3}} and galactic cluster emissions of X-rays {\cite{c4}}
all convincingly indicate that the observable universe is
undergoing an accelerated expansion. The data also show that this
sudden change in the expansion of the universe is quite recent
($z\simeq0.7$) while previously the universe was in the phase of
deceleration. The origin of acceleration may be caused due to
exotic `dark energy' (DE) in the universe. It is now the job of
experimentalists to parameterize DE and for theorists to determine
its origin. Although the simplest way to explain this behavior is
the consideration of a cosmological constant \cite{c7}, the known
fine-tuning problem (huge discrepancies in the theoretical and
observational estimates of DE parameters particularly in its
energy density) and coincidence problem (apparent constant ratio
of energy densities of DE and matter) \cite{8} led to the DE
paradigm. The dynamical nature of DE, at least in an effective
level, can arise from a variable cosmological ``constant''
\cite{varcc}, or from various fields, such is a canonical scalar
field (quintessence) \cite{quint}, a phantom field, that is a
scalar field with a negative sign of the kinetic term
\cite{phant}, or the combination of quintessence and phantom in a
unified model named quintom \cite{quintom}. Finally, an
interesting attempt to probe the nature of DE according to some
basic quantum gravitational principles is the holographic DE (HDE)
paradigm \cite{holoext} which we shall discuss in ample detail
below.

Recent studies of black hole and string theories provide an
alternative solution to the DE puzzle using the holographic
principle \cite{cohen} originally proposed by t' Hooft
\cite{hooft}. The principle says that the entropy of a closed
system scales not by its volume but with its surface area. In
other words the degrees of freedom of a spatial region reside not
in the bulk but only at the boundary of the region and the number
of degrees of freedom per Planck area are not greater than unity.
It is widely accepted that this principle will be a part of the
final quantum gravity since both string theory and loop quantum
gravity incorporate this principle beautifully.

Cohen et al. \cite{cohen} suggested that in quantum field theory a
short distance cutoff is related to a long distance cutoff due to
the limit set by formation of a black hole, which results in an
upper bound on the zero-point energy density. In line with this
suggestion, some authors \cite{holoext} argued that this energy
density could be viewed as the HDE density satisfying $\rho_{\Lambda
} =3n^2M^2_p/L^2$, where $L$ is the size of a region which provides
an IR cut-off, and the numerical constant $3n^2$ is introduced for
convenience. It is important to note that in the literature, various
scenarios of HDE have been studied via considering different
system's IR cutoff \cite{jamil1}. If one takes the Hubble length as
the IR cut-off ($\rho_D\propto H^2$) then it conveniently resolves
the fine tuning problem but yields a wrong equation of state of DE
($w_D=0$) which cannot drive cosmic acceleration. Moreover a
different IR cut-off, a particle horizon, also yields a wrong
equation of state ($w_D>-1/3$) of DE. Later on Granda and Oliveros
\cite{GO} proposed a new cut-off based on purely dimensional
grounds, by adding a term involving the first derivative of the
Hubble parameter. Thus the new cut-off is similar to the Ricci
scalar of the FRW metric $\rho_D\sim\gamma H^2+\beta \dot H$, where
$\gamma$ and $\beta$ are constants of order unity. It was predicted
that their values should be $\gamma\simeq0.93$ and $\beta\simeq0.5$,
in order to be consistent with the big bang nucleosynthesis theory
\cite{GO}. Other studies on the HDE have been carried out in
\cite{Setare1}.

Soon after Albert Einstein introduced his ``general theory of
relativity" in $1915$, several attempts were made to construct
alternative theories of gravity. These were intended to construct
unified models of all forces. One of the most studied alternative
theories was scalar-tensor theory, where the gravitational action
contains, apart from the metric, a scalar field which describes part
of the gravitational field. The scalar-tensor theory was invented
first by Jordan \cite{Jor} in the 1950's, and then taken over by
Brans and Dicke \cite{BD} some years later. Soon after the discovery
of dark matter (DM) and DE, a new breed of gravities like Gauss
Bonnet and f(R) gravities proposed whose Lagrangian contained
several terms involving curvature tensor and scalars. Because the
HDE density belongs to a dynamical cosmological constant, we need a
dynamical frame to accommodate it instead of Einstein gravity.
Therefore the investigation on the holographic models of DE in the
framework of Brans-Dicke theory is of great importance
\cite{Setare2}. The investigation on the holographic models of DE in
the framework of Brans-Dicke cosmology, have been carried out in
\cite{other,Pavon2,Sheykhi1,Bert,Xu,jamil}.

In this paper, we investigate the HDE in the Brans-Dicke gravity
using the Granda-Oliveros cut-off. We follow the method of Ref.
\cite{Setare2}. Following previous studies, we calculate the
equation of state of DE and some other cosmological parameters of
our interest and demonstrate that phantom crossing is possible in
the present model.

\section{New HDE in Brans-Dicke gravity}

The action of Brans-Dicke theory is \cite{xin}
\begin{equation}\label{action}
S=\int d^4x\sqrt{g}\Big[ \frac{1}{2}\Big( \Phi
R-\omega\frac{\nabla_\mu\Phi\nabla^\mu\Phi}{\Phi} \Big)+L_m \Big].
\end{equation}
The equations of motion for the metric $g_{\mu\nu}$ and the
Brans-Dicke scalar field $\Phi$ are
\begin{eqnarray}\label{efe}
R_{\mu\nu}-\frac{1}{2}g_{\mu\nu}R&=&\frac{1}{\Phi}T_{\mu\nu}^M+T_{\mu\nu}^{BD},\\
\nabla_\mu\nabla^\mu\Phi&=&\frac{1}{2\omega+3}T^{M\mu}_\mu.
\end{eqnarray}
Here $T_{\mu\nu}^M=(2/\sqrt{g})\delta(\sqrt{g}L_M)/\delta
g^{\mu\nu}$ is the energy momentum tensor for the matter fields
defined in the form of perfect fluid
\begin{equation}\label{tm}
T_{\mu\nu}^M=(\rho_M+p_M)U_\mu U_\nu+p_Mg_{\mu\nu}.
\end{equation}
The energy momentum tensor of the Brans-Dicke scalar field is
\begin{equation}\label{tbd}
T_{\mu\nu}^{BD}=\frac{\omega}{\Phi^2}\Big(
\nabla_\mu\Phi\nabla_\nu\Phi-\frac{1}{2}g_{\mu\nu}\nabla_\alpha \Phi\nabla^\alpha\Phi
\Big)+\frac{1}{\Phi}(\nabla_\mu\nabla_\nu\Phi-g_{\mu\nu}\nabla_\alpha\nabla^\alpha\Phi).
\end{equation}
The governing equations of the system are
\begin{equation}\label{first}
H^2+\frac{k}{a^2}+H\frac{\dot
\Phi}{\Phi}-\frac{\omega}{6}\Big(\frac{\dot
\Phi}{\Phi}\Big)^2=\frac{1}{3\Phi}(\rho_M+\rho_D),
\end{equation}
\begin{equation}\label{second}
2\frac{\ddot a}{a}+H^2+\frac{k}{a^2}+\frac{\omega}{2}\Big(\frac{\dot
\Phi}{\Phi}\Big)^2+2H\frac{\dot
\Phi}{\Phi}+\frac{\ddot\Phi}{\Phi}=-\frac{p_D}{\Phi}.
\end{equation}
The energy conservation equations for DE and matter are
\begin{equation}\label{drho}
\dot{\rho}_D+3H(1+w_D)\rho_D=0,
\end{equation}
\begin{equation}\label{dm}
\dot{\rho}_m+3H\rho_m=0.
\end{equation}
We use the new HDE \cite{GO} which is modified to be consistent with
the BD framework
\begin{equation}\label{hde}
\rho_D=3\Phi(\gamma H^2+\beta\dot H),
\end{equation}
where $\gamma$ and $\beta$ are constants. Notice that the above cut
off (\ref{hde}) is different from the Ricci dark energy \cite{ricci}
and reduces to the later one if $k=0$, $\beta=1$ and $\gamma=2$.  We
choose the ansatz
\begin{equation}\label{phi}
\Phi(a)=\Phi_0a(t)^\alpha,\ \ \dot\Phi=\alpha\Phi H,\ \
\ddot\Phi=\Phi(\alpha^2H^2+\alpha\dot H),
\end{equation}
where in what follows we take $\Phi_0=1$. Note that the system of
Eqs. (6)-(7) is not closed and we still have freedom to choose one.
We shall assume that Brans-Dicke field can be described as a power
law of the scale factor, $\phi\propto a^{\alpha}$. In principle
there is no compelling reason for this choice. However, it has been
shown that for small $\alpha$ it leads to consistent results
\cite{Pavon2,Xu}. A case of particular interest is that when
$\alpha$ is small whereas $\omega$ is high so that the product
$\alpha\omega$ results of order unity \cite{Pavon2}. This is
interesting because local astronomical experiments set a very high
lower bound on $\omega$; in particular, the Cassini experiment
implies that $\omega>10^4$ \cite{Bert}. Introducing a new parameter
$x=\ln a$, the e-folding parameter. Using it we get
\begin{equation}\label{hdot}
\dot H=\frac{1}{2}\frac{dH^2}{dx}.
\end{equation}
Assuming matter evolves as $\rho_m=\rho_{m_0}a^{-3}$. Using
(\ref{hde}), (\ref{phi}) and (\ref{hdot}) in (\ref{first}), we
obtain
\begin{equation}\label{dh2}
\frac{dH^2}{dx}-\frac{2\delta}{\beta}H^2=-\frac{2\rho_{m_0}}{3\beta}e^{-x(\alpha+3)}+\frac{2k}{\beta}e^{-2x},
\end{equation}
where
\begin{equation}
\delta\equiv 1-\gamma+\alpha\Big(1-\frac{\alpha\omega}{6}\Big).
\end{equation}
Equation (\ref{dh2}) can be solved to get
\begin{equation}\label{h2}
H^2=\frac{2\rho_{m_0}}{3(\alpha\beta+2\delta+3\beta)}e^{-x(\alpha+3)}+c_1e^{2\delta
x/\beta}-\frac{k}{\beta+\delta}e^{-2x},
\end{equation}
where $c_1$ is a constant of integration. Using (\ref{hde}) in
(\ref{drho}), we obtain
\begin{equation}\label{omd}
w_D=-1-\frac{\alpha}{3}-\frac{2\gamma H\dot H+\beta\ddot H}{3H(\gamma H^2+\beta\dot
H)}.
\end{equation}
Using (\ref{h2}) in (\ref{omd}), we get the equation of state (EoS)
parameter for the new HDE in the framework of the Brans-Dicke theory
\begin{equation}\label{od}
w_D=-1-\frac{\alpha}{3}-\frac{\frac{[\beta(\alpha+3)-2\gamma](\alpha+3)}{3(\alpha\beta+2\delta+3\beta)}\rho_{m_0}e^{-x(\alpha+3)}
+2c_1(\frac{\delta}{\beta})(\gamma+\delta)e^{2\delta
x/\beta}+2\frac{k}{\beta+\delta}(\gamma-\beta)e^{-2x}}
{\frac{-[\beta(\alpha+3)-2\gamma]}{(\alpha\beta+2\delta+3\beta)}\rho_{m_0}e^{-x(\alpha+3)}+3c_1(\gamma+\delta)e^{2\delta
x/\beta}-3\frac{k}{\beta+\delta}(\gamma-\beta)e^{-2x}},
\end{equation}
or
\begin{equation}\label{od1}
w_D=-1-\frac{\frac{[\beta(\alpha+3)-2\gamma]}{(\alpha\beta+2\delta+3\beta)}\rho_{m_0}e^{-x(\alpha+3)}
+c_1(\alpha+\frac{2\delta}{\beta})(\gamma+\delta)e^{2\delta
x/\beta}-\frac{k}{\beta+\delta}(\gamma-\beta)(\alpha-2)e^{-2x}}
{\frac{-[\beta(\alpha+3)-2\gamma]}{(\alpha\beta+2\delta+3\beta)}\rho_{m_0}e^{-x(\alpha+3)}+3c_1(\gamma+\delta)e^{2\delta
x/\beta}-3\frac{k}{\beta+\delta}(\gamma-\beta)e^{-2x}}.
\end{equation}
In the limiting case $\alpha=0$ ($\omega\rightarrow\infty$), the
Brans-Dicke scalar field becomes trivial and Eq. (\ref{od1}) in the
absence of matter ($\rho_{m_0}\rightarrow 0$) reduces to its
respective expression in new HDE model in Einstein gravity
\cite{Karami2}
\begin{equation}\label{od21}
w_D=-\frac{1}{3}\left(\frac{k\big(\frac{\gamma-\beta}{\beta+\delta}\big)e^{-2x}-c_1(3+\frac{2\delta}{\beta})(\gamma+\delta)e^{2\delta
x/\beta}}
{{k\big(\frac{\gamma-\beta}{\beta+\delta}\big)e^{-2x}-c_1(\gamma+\delta)e^{2\delta
x/\beta}}}\right).
\end{equation}
If we compare Eq. (\ref{od}) with Eq. (\ref{od21}) we find out that
when the new HDE is combined with Brans-Dicke field the transition
from normal state where $w_D>-1 $ to the phantom regime where
$w_D<-1$ for the EoS of DE \cite{Wang} can be more easily achieved
for than when resort to the Einstein field equations is made. The
analysis of the properties of DE from recent observations mildly
favor models with $\omega_{D}$ crossing -1 in the near past
\cite{Alam}.

To illustrate this result in ample detail, we investigate it for the
late-time universe when $x\rightarrow \infty$. In this case, Eq.
(17) reduces to
\begin{equation}
w_D=-1-\frac{2}{3}\Big(\frac{1-\gamma}{\beta}\Big)
-\frac{\alpha}{3}\Big[1+\frac{2}{\beta}\Big(1-\frac{\alpha\omega}{6}\Big)\Big],\label{wBDlate}
\end{equation}
and Eq. (19) yields
\begin{equation}
w_D=-1-\frac{2}{3}\Big(\frac{1-\gamma}{\beta}\Big).\label{wGRlate}
\end{equation}
If we take $\alpha\omega\approx1$ \cite{Pavon2}, $\gamma\simeq0.93$
and $\beta\simeq0.5$ \cite{GO} then for the new HDE in Brans-Dicke
gravity, Eq. (\ref{wBDlate}) gives $w_D=-1.09-1.44\alpha$ and in
Einstein gravity ($\alpha\rightarrow0$) from Eq. (\ref{wGRlate}) we
obtain $w_D=-1.09$. Thus in the late-time universe, crossing the
phantom divide line for the new HDE in Brans-Dicke gravity can be
more easily achieved for than when resort to the Einstein gravity.

Using (\ref{second}), the deceleration parameter can be evaluated
to be
\begin{equation}\label{q}
q=-\frac{\ddot a}{aH^2}=\frac{1}{(2+\alpha+3\beta w_D )}\Big[
(1+\alpha)^2+\frac{\omega}{2}\alpha^2-\alpha+3(\gamma-\beta)w_D
+\Omega_k\Big].
\end{equation}
The evolution of dimensionless DE parameter is
\begin{equation}\label{Op}
\Omega'_D=-3\gamma\Big(w_D+1+\frac{\alpha}{3}\Big)+\Big(\frac{\gamma-\Omega_D}{\beta}\Big)\Big[\beta(3w_D+3+\alpha)+2\gamma\Big]-
2\beta\Big(\frac{\gamma-\Omega_D}{\beta}\Big)^{2}.
\end{equation}

\section{Interacting new HDE in Brans-Dicke gravity}
Next we generalize our study to the case of interacting new HDE in
Brans-Dicke theory. Recent cosmological observations support the
interaction between DE and DM \cite{Bertolami8}. Taking the
interaction into account, HDE and DM do not conserve separately
and enter the energy balances \cite{Zimdahl}
\begin{eqnarray}
&&\dot{\rho}_D+3H\rho_D(1+w_D)=-Q,\label{consq2}\\
&&\dot{\rho}_{DM}+3H\rho_{DM}=Q, \label{consm2}\\
&&\dot{\rho}_{BM}+3H\rho_{BM}=0,\label{consbm}
\end{eqnarray}
where we have assumed the BM dose not interact with DE. Here Q
denotes the interaction term and we take it as
\begin{eqnarray}
Q = 3b^2H(\rho_{DM}+\rho_D),
\end{eqnarray}
where $b^2$ is a coupling constant. The critical energy density,
$\rho_{\mathrm{cr}}$, and the energy density of the curvature,
$\rho_k$, are defined as
\begin{eqnarray}\label{rhocr}
\rho_{\mathrm{cr}}=3\Phi H^2,~~~~~~~\rho_k=\frac{3k\Phi}{a^2}.
\end{eqnarray}
We also introduce the fractional energy densities such as
\begin{eqnarray}
\Omega_{BM}&=&\frac{\rho_{BM}}{\rho_{\mathrm{cr}}}=\frac{\rho_{BM}}{3\Phi
H^2}, \label{Omegam} \\
\Omega_{DM}&=&\frac{\rho_{DM}}{\rho_{\mathrm{cr}}}=\frac{\rho_{DM}}{3\Phi
H^2}, \label{Omegam} \\
\Omega_D&=&\frac{\rho_D}{\rho_{\mathrm{cr}}}=\frac{\rho_{D}}{3\Phi
H^2},\\
\Omega_{k}&=&\frac{\rho_{k}}{\rho_{\mathrm{cr}}}=\frac{k}{a^2
H^2}.\label{Omegam}\label{OmegaD}
\end{eqnarray}
Combining Eqs. (\ref{rhocr}) and (\ref{phi}) with the first
Friedmann equation (\ref{first}), we can rewrite this equation as
\begin{eqnarray}\label{rhos}
\rho_{\mathrm{cr}}+\rho_k=\rho_{BM}+\rho_{DM}+\rho_D+\rho_{\Phi},
\end{eqnarray}
where we have defined
\begin{eqnarray}\label{rhophi}
\rho_{\Phi}\equiv \alpha
H^2\Phi\left(\frac{\alpha\omega}{2}-3\right).
\end{eqnarray}
Dividing Eq. (\ref{rhos}) by $\rho_{\rm cr}$, it can be rewritten
as
\begin{eqnarray}\label{Fried2new}
\Omega_{BM}+\Omega_{DM}+\Omega_D+\Omega_{\Phi}=1+\Omega_k,
\end{eqnarray}
where
\begin{eqnarray}\label{Omegaphi}
\Omega_{\Phi}=\frac{\rho_{\Phi}}{\rho_{\mathrm{cr}}}=-\frac{\alpha}{3}
\left(3-\frac{\alpha \omega}{2}\right).
\end{eqnarray}
Therefore  we can rewrite the interaction term as
\begin{eqnarray}\label{Q}
Q=3b^2H(\rho_{DM}+\rho_D)=3b^2H\rho_D(1+u),
\end{eqnarray}
where
\begin{eqnarray}\label{r}
u=\frac{\rho_{DM}}{\rho_D}=-1+\frac{1}{\Omega_D}\left[1+\Omega_k-\Omega_{BM}
+\frac{\alpha}{3} \left(3-\frac{\alpha \omega}{2}\right)\right],
\end{eqnarray}
is the energy density ratio of two dark components.

Following the approach of the previous section we obtain the EoS
parameter of the interacting new HDE in Brans-Dicke theory as
\begin{eqnarray}
w_D=-1-\frac{\frac{[\beta(\alpha+3)-2\gamma]}{(\alpha\beta+2\delta+3\beta)}\rho_{m_0}e^{-x(\alpha+3)}
+c_1(\alpha+\frac{2\delta}{\beta})(\gamma+\delta)e^{2\delta
x/\beta}-\frac{k}{\beta+\delta}(\gamma-\beta)(\alpha-2)e^{-2x}}
{\frac{-[\beta(\alpha+3)-2\gamma]}{(\alpha\beta+2\delta+3\beta)}\rho_{m_0}e^{-x(\alpha+3)}+3c_1(\gamma+\delta)e^{2\delta
x/\beta}-3\frac{k}{\beta+\delta}(\gamma-\beta)e^{-2x}}
\nonumber\\
-\frac{b^2}{\Omega_D}\left[1+\Omega_k-\Omega_{BM}
+\frac{\alpha}{3} \left(3-\frac{\alpha
\omega}{2}\right)\right].\label{wint}
\end{eqnarray}
Comparing Eq. (39) with (17) shows that in the presence of
interaction since the last expression in Eq. (39) has a negative
contribution, hence crossing the phantom divide, i.e. $w_D<-1$, can
be more easily achieved for than when the interaction between the
new HDE and DM is not considered.

The deceleration parameter $q$ and the equation of motion for
$\Omega_{D}$ are still obtained according to Eqs. (\ref{q}) and
(\ref{Op}), respectively, where $w_{D}$ is now given by Eq.
(\ref{wint}).

\section{Concluding remarks}

We discuss the role of the HDE with the Granda-Oliveros cut-off in
the Brans-Dicke theory. It is very interesting to investigate the
role of dynamical cosmological constant (HDE) in the dynamical
framework (Brans-Dicke theory). The Granda-Oliveros length scale
is a natural extension of the Hubble length since the later does
not resolve some DE issues such as the equation of state of DE. We
have found that the new equation of state obtained above provides
necessary corrections and enables a phantom crossing/divide of the
state parameter. We also consider an interaction between DE and DM
(ignoring the baryonic component) and found that phantom crossing
is softer in this case as compared to the non-interacting case.

\acknowledgments{ The works of K. Karami and A. Sheykhi have been
supported financially by Research Institute for Astronomy and
Astrophysics of Maragha, Iran.}



\end{document}